\documentclass[8pt,prd,aps,showpacs,nofootinbib,preprint,eqsecnum]{revtex4-1}
\usepackage{hyperref}
\usepackage{latexsym}
\usepackage{epstopdf}
\usepackage{epsf}
\usepackage{amssymb,amsmath,amsfonts,amsxtra,amsthm}
\usepackage{bm}
\usepackage{dcolumn}
\usepackage{array}
\usepackage{tabularx}
\usepackage{enumerate}
\usepackage{hhline}
\usepackage{subfigure}
\usepackage{color}
\usepackage{ulem}
\newcommand{\beq}{\begin{equation}}
\newcommand{\eeq}{\end{equation}}
\newcommand{\beqa}{\begin{eqnarray}}
\newcommand{\eeqa}{\end{eqnarray}}
\newcommand{\barr}{\begin{array}}
\newcommand{\earr}{\end{array}}
\newcommand{\bib}[1]{\bibitem{#1}}
\begin{document}

\title{Primordial Fluctuations within Teleparallelism}
\date{\today}

\author{Yi-Peng Wu$^{1, 3}$\footnote{E-mail address: 
s9822508@m98.nthu.edu.tw} and 
Chao-Qiang Geng$^{1,2,3}$\footnote{E-mail address: geng@phys.nthu.edu.tw} 
}
\affiliation{
$^1$Department of Physics, National Tsing Hua University, Hsinchu, Taiwan 300 
\\ 
$^2$Physics Division, National Center for Theoretical Sciences, Hsinchu, Taiwan 300
\\
$^3$Kavli Institute for Theoretical Physics China (KITPC), 
CAS,
Beijing 100190, China
}
\begin{abstract}

We study the cosmological perturbations for the possible inflation scenario in the 
teleparallel equivalence of general relativity specified with parallelizable topological conditions. 
By acquiring the identical physical observables to general relativity under the teleparallel formalism, 
we perform a 3+1 decomposition of the vierbein field, which can be interpreted as the time gauge
fixing between coordinate and tangent frames. 
We also extend our discussion to the higher-order action, $f(T)$ gravity.
\end{abstract}

\keywords{$f(T)$ theory}

\maketitle
\section{Introduction}

Inflation is one of the most successful cosmological models, 
which resolves problems of the flatness, the horizon, 
and provides explanation to the structure of the universe observed today. 
The physical observables of each inflationary model are obtained through the quantization of the primordial fluctuations. Such scenario is demonstrated in~\cite{TMuk,nonMal} for the simplest case that gravity is minimally coupled to a scalar field. The discussion can be also applied to higher-order gravity actions which have conformal equivalences to the minimal coupling cases~\cite{TMuk}. More generically, it can be used in the most general second-order field equations~\cite{GG,GI} or  general modified gravitational models of inflation~\cite{GMG,Gao}. The problem of inflation can be also attacked via bifurcation arguments~\cite{Inf_bi}.

Recently, the gravity constructed within ``distant parallelism'' or ``teleparallelism'' has received much attention of cosmological interests for both  
the early universe~\cite{iwi} and  the late time cosmic 
acceleration~\cite{ft1,ft2,ft3,ft4,ft4,ft5,ft6,ft7,ft9,ft10,ft11,ft12,ft13,ft14,ft15,ft16,ft17,ft18,ft19,ft20,ft21,ft22,ft23,ft24,LV,dof,CT,LSS}. 
While the teleparallel equivalence of general relativity (TEGR)~\cite{E} is, so far, indistinguishable from general relativity (GR), its high order generalization, 
$f(T)$ gravity, contains some novel features other than $f(R)$ gravity. 
Remarkably, the recent development of teleparallel gravity~\cite{iwi,ft1,ft2,ft3,ft4,ft4,ft5,ft6,ft7,ft9,ft10,ft11,ft12,ft13,
ft14,ft15,ft16,ft17,ft18,ft19,ft20,ft21,ft22,ft23,ft24,LV,dof,CT,LSS} follows
the geometric construction in~\cite{NGR,T intro}, where a
special affine connection with absolute parallelism is used.
This construction imposes a strong topological condition in teleparallelism,
interpreted as the vanishing connection coefficients (or spin-connection), 
 ensuring the independence of parallel transformations between 
the four orthonormal bases of the vector field~\cite{gravity and gauge symm}.
Although, in general, the teleparallel condition can be realized through a
constraint to the curvature tensor~\cite{Hehl},
the formalism in this work represents a technically simpler approach to
acquire physical observables from the dynamical vierbein field and
 suffices for our interest of cosmological perturbations.

Nevertheless, it is noticeable that such topologically
``parallelizable'' condition inevitably breaks the local Lorentz invariance  
while some extra degrees of freedoms are found~\cite{LV,dof}. 
This feature can be explicit in the Einstein frame of $f(T)$ gravity 
 as an exotic scalar-torsion coupling appears in the Lagrangian~\cite{CT}. 
 However, under homogeneous and isotropic principles in cosmology, 
 the effects of the extra degrees of freedom  proceed to the perturbed equations. It is shown
 in~\cite{LSS} that under a gauge-invariant analysis, one extra degree of freedoms causes severe constraints on $f(T)$ gravity. 

 In this study, we investigate the cosmological perturbations
of some possible inflationary models under the teleparallel description of GR. 
A 3+1 decomposition, specifying the vierbein fields, can lead to 
the ADM formulation with the metric commonly used in the literature.  
After applying such vierbein fields to TEGR, we find that the resulting 
torsion scalar $T$ is identical to the Hamiltonian formulation under 
a certain time gauge consideration~\cite{HT}. 
This time gauge gives rise to the class of the 
teleparallel geometry with the definite identification on the energy and 
momentum of the gravitational field, and meanwhile recovers the local Lorentz
invariance of the theory. The representation is also convenient 
for comparison between the formulations of GR.
For the general torsion framework including field equations, one can refer to~\cite{Hehl}. 

In this paper, we demonstrate that TEGR shares the same quadratic actions as GR for 
both scalar and tensor perturbations, which ensure that all the physical 
observables are unchanged by choosing the special affine connection with absolute parallelism.
We also discuss the possible inflation scenario driven purely by the higher-order gravity in 
teleparallelism as a similar \textit{ad hoc} to $f(R)$ theories~\cite{A.A}. 
With regard to the similar time gauge condition to the vierbein fields, 
$f(T)$ theories provide to be as special cases of the general 
density perturbations with second order field equations~\cite{GG}.
This result gives a preliminary study of the cosmological perturbations in
teleparallel theories fixed with local Lorentz invariance,
while further investigation on extra degrees of freedom 
introduced by the parallelizable condition requires for consideration beyond 
metric anstaz. 
For the \textsl{coordinate and tangent frames}, Greek  indices $\mu, \nu,...$ and 
capital Latin indices $A,B,...$ run over space and time, while Latin indices $i,j,...$ and $a,b,...$, 
represent the spatial part of 1, 2, 3, respectively.

The paper is organized as follows. 
In Sec.~II, we review  teleparallel gravity. 
In Sec.~ III, we perform the quadratic computation in TEGR.
 We extend our discussion to the high order generalization of TEGR in Sec.~IV.
Finally, conclusions are given in Sec.~V.

\section{Teleparallel gravity}

Originated from the Einstein's approach to unify gravitation with electromagnetism \cite{E}, 
teleparallel gravity takes the vierbein fields
$\mathbf{e}_A(x^\mu)$ as the dynamical variables, which are also orthonormal bases of the tangent space for each point
$x^\mu$\, on the manifold. These bases satisfy the relation
$\mathbf{e}_A\cdot\mathbf{e}_B=\eta_{AB}$, where
$\eta_{AB}=\mbox{diag}(1,-1,-1,-1)$, and are related to the coordinate bases
$\partial_\mu$ via the components\;$e^\mu_A$, $i.e.$, $\mathbf{e}_A=e^\mu_A\partial_\mu$. Therefore, the metric tensor is obtained from the dual vierbein as
\beq \label{metric}
g_{\mu\nu}(x)=\eta_{AB}\,e^A_\mu(x) e^B_\nu(x).
\eeq

If the metric compatible condition $\nabla_\rho g_{\mu\nu}=0$ is
generalized to the vierbein fields, it will lead to a vanishing total covariant derivative
of $e^A_\mu$, given by~\cite{Ortin}
\beq \label{TD}
D_\mu e^A{}_\nu=\partial_\mu e^A{}_\nu
             -\Gamma^{\rho}_{\,\,\nu\mu}e^A{}_\rho
             +\omega^{A}_{\,\,B\mu} e^B{}_\nu=0,
\eeq
where $\Gamma^{\rho}_{\,\,\nu\mu}$ is the affine connection of the coordinate covariant 
derivative $\nabla_\mu$, while the spin-connection $\omega^{A}_{\,\,B\mu}$ describes the
same geometric object as $\Gamma$ in the tangent frame. 
By requiring a quadruplet of the linearly independent orthonormal basis $\mathbf{e}_A$, 
 teleparallel gravity~\cite{NGR} 
uses the curvatureless Weitzenb\"{o}ck connection, 
$\Gamma^{\rho}_{\,\,\mu\nu}=e^\rho_A\partial_\nu e^A_\mu$, to define its covariant derivative.
We can see from (\ref{TD}) that this special form of  the connection indicates
the ``parallelizable'' condition, $\omega^{A}_{\,\,B\mu}=0$, of 
the spacetime, and  breaks the local Lorentz invariance in the tangent frame~\cite{T intro,LV},
while the teleparallel condition is satisfied identically due to the vanishing spin-connection since
\beq \nonumber
\mathcal{R}^A{}_{B\nu\mu}=
\partial_\nu\omega^{A}_{\,\,B\mu}-\partial_\mu\omega^{A}_{\,\,B\nu}+
\omega^{A}_{\,\,C\nu}\omega^{C}_{\,\,B\mu}-\omega^{A}_{\,\,C\mu}\omega^{C}_{\,\,B\mu},
\eeq 
where $\mathcal{R}^\lambda{}_{\rho\nu\mu}\equiv e_A^\lambda e_\rho^B\mathcal{R}^A{}_{B\nu\mu}$ 
is the curvature tensor. 
It is noteworthy that teleparallel gravity is now completely determined by the vierbein fields.
Although, in principle, TEGR does not have to be treated in such a Lorentz violation formulation, 
the condition $\omega^{A}_{\,\,B\mu}=0$ can practically simplify the calculations without 
affecting the physical observables of the theory~\cite{gravity and gauge symm}.
The explicit examination will be provided in the next section.

Following the Weitzenb\"{o}ck connection
$\Gamma^{\rho}_{\,\,\mu\nu}=e^\rho_A\partial_\nu e^A_\mu$, 
the torsion tensor is given by
\beq
T^{\rho}_{\,\,\mu\nu}=\Gamma^{\rho}_{\,\,\nu\mu}-\Gamma^{\rho}_{\,\,\mu\nu}= e^\rho_A(\partial_\mu e^A_\nu-\partial_\nu e^A_\mu).
\eeq
The relation between the Weitzenb\"{o}ck connection and the torsionless Levi-Civita connection $\bar{\Gamma}^{\rho}_{\,\,\mu\nu}$ used in GR is given through the contorsion tensor $K^{\rho}_{\,\,\mu\nu}=\frac{1}{2}(T^{\,\,\rho}_{\nu\;\,\mu} +T^{\,\,\rho}_{\mu\;\,\nu}-T^{\rho}_{\,\,\mu\nu})$ as
\beq \label{w-l}
\Gamma^{\rho}_{\,\,\mu\nu}=\bar{\Gamma}^{\rho}_{\,\,\mu\nu}+K^{\rho}_{\,\,\mu\nu}.
\eeq

As demonstrated in \cite{NGR}, the most general Lagrangian density, which is the quadratic of the torsion tensor, is of the form
\beq \label{lngr}
\mathcal{L}=a_1 T_{\rho}^{\,\,\mu\nu}T^{\rho}_{\,\,\mu\nu}+a_2 T^{\mu\nu}_{\,\,\,\,\,\,\,\,\rho} T^{\rho}_{\,\,\mu\nu}+a_3 T^{\rho}_{\,\,\rho\mu}T^{\nu\,\,\mu}_{\,\,\nu}+a_0,
\eeq
where $a_i$ are free parameters. An equivalent description to GR within  teleparallelism is established in \cite{HT} and can be simply taken as the Lagrangian (\ref{lngr}) with the choice
$a_1=1/8, a_2=-1/4$ and $a_3=-1/2$ \cite{T intro}.
Consequently, the Lagrangian density of 
TEGR is given by
\beqa
\label{TEGRL}
\mathcal{L}_{TEGR}=\frac{1}{2}T&=&\frac{1}{8}T_{\rho}^{\,\,\mu\nu}T^{\rho}_{\,\,\mu\nu}-\frac{1}{4}T^{\mu\nu}_{\,\,\,\,\,\,\,\,\rho} T^{\rho}_{\,\,\mu\nu}-\frac{1}{2}T^{\rho}_{\,\,\rho\mu}T^{\nu\,\,\mu}_{\,\,\nu}
\nonumber \\
                               &\equiv &\frac{1}{2}S_\rho^{\;\,\mu\nu}T^{\rho}_{\,\,\mu\nu},
\eeqa 
with the definition 
$S_\rho^{\;\,\mu\nu}=\frac{1}{2}(K_{\;\;\;\rho}^{\mu\nu}+\delta^\mu_\rho T^{\alpha\nu}_{\;\;\;\alpha}-\delta^\nu_\rho T^{\alpha\mu}_{\;\;\;\alpha})$. The field equation of TEGR is obtained by variation of the vierbein field $e_{A}^{\mu}$ with respect to the action 
$S=\frac{1}{2}\int d^4x[eT+\mathcal{L}_m]$, 
 given by
\begin{eqnarray}\label{eom}
e^{-1}\partial_{\mu}(ee_{A}^{\rho}S_{\rho}{}^{\mu\nu})
-e_{A}^{\lambda}T^{\rho}{}_{\mu\lambda}S_{\rho}{}^{\nu\mu}
-\frac{1}{4}e_{A}^{\nu}T
= \frac{1}{2}e_{A}^{\rho}\Theta_{\rho}{}^{\nu},
\end{eqnarray}
where $\Theta_{\rho}{}^{\nu}\equiv e^A_\rho e^{-1}\delta\mathcal{L}_m/\delta e^A_\nu$ 
is the energy-momentum tensor of the matter. 
This energy-momentum tensor has to be symmetric,
$\Theta_{\mu\nu}=\Theta_{\nu\mu}$, and conserved with respect to the Levi-Civita
covariant derivative, $\bar{\nabla}^\mu\Theta_{\mu\nu}=0$, as 
guaranteed by the invariant action principle of the general coordinate 
and local Lorentz transformations of the matter sector, $S_m=\int d^4x\mathcal{L}_m$~\cite{LV,Weinberg}.
By using the relation (\ref{w-l}), it is found that 
$G_{A}^\nu=2e^{-1}\partial_{\mu}(ee_{A}^{\rho}S_{\rho}{}^{\mu\nu})
-2e_{A}^{\lambda}T^{\rho}{}_{\mu\lambda}S_{\rho}{}^{\nu\mu}
-\frac{1}{2}e_{A}^{\nu}T$, 
where $G_{\mu\nu}=e^A_{\;\mu}G_{A}^{\;\nu}$ is nothing but the Einstein tensor. 
Therefore, we can rewrite (\ref{eom}) in the covariant way as $G_{\mu\nu}=\Theta_{\mu\nu}$, which is completely the same geometrical formulation as the Einstein equation in GR.
%

\section{Quadratic Computation}

The computation of the second order action for the primordial fluctuations of the standard inflationary model
\beq \label{LR}
S=\frac{1}{2}\int d^4x\sqrt{-g}[R+(\nabla\phi)^2-2V(\phi)]
\eeq
has been reviewed nicely in \cite{TMuk} as well as \cite{nonMal} with a scenario commonly used in modified gravity theories which we shall follow in the rest of our discussion. 
Note that $R$ is the Ricci scalar of GR defined 
purely by the metric tensor (\ref{metric}).

In order to make a clear comparison to the standard results, we consider the existence of a scalar field with a general potential
in the Lagrangian of TEGR, given by
\beq \label{LT}
S=\frac{1}{2}\int d^4x e[T+(\nabla\phi)^2-2V(\phi)],
\eeq
where $e\equiv\sqrt{-g}$ and the torsion scalar $T$ is defined in (\ref{TEGRL}).
 From the mathematical manipulation by using the relation between Weitzenb\"{o}ck and Levi-Civita connections, one gets~\cite{T intro}
\begin{eqnarray}
\label{T}
T&=&
 R+2\bar{\nabla}^{\nu}T^{\mu}_{\,\,\mu\nu}\,,
\end{eqnarray}
which shows that $T$ and $R$ only differ by a divergent term 
with respect to the Levi-Civita derivative.
It is clear that the divergent term in (\ref{T}) is not a metrical quantity,
which does not respect to the local Lorentz invariance of the tangent frame~\cite{LV,T intro}.
However,  such term appears as a total derivative in the
action so that it does not contribute to the field equations. 

When we choose the vierbein to be $e^A_{\mu}=\mbox{diag}(1,a,a,a)$,  (\ref{LT}) shares the same background equations as (\ref{LR}) in the flat FRW background:
\begin{eqnarray}
\label{bgeq}
3H^2=\frac{1}{2}\dot{\phi}^2+V\,,\
\dot{H}=-\frac{1}{2}\dot{\phi}^2\,,\ and\,\;\
0=\ddot{\phi}+3H\dot{\phi}+V_{,\phi}.
\end{eqnarray}
Note that such a simple vierbein choice is an exact solution of the TEGR field equation corresponding to the flat FRW Einstein equation followed by the discussion in \cite{NGR}, and one can refer to \cite{ft14} for the cases of open and closed universe.  

\subsection{ADM Decomposition}

The computation of the quadratic action is commonly proceeded in the 
ADM formalism for various inflationary theories to achieve the nearly scale 
invariant curvature perturbations. 
For torsion theories, the ADM type formulation has been 
studied in \cite{Ann phys, gravity and gauge symm, LE GR}.
The ADM decomposition of the \textsl{coordinate},
despite specifying the metric field, does not fix all the components of the vierbein.
In the present work, we consider  a specific vierbein choice, given by
\beqa \label{ADMvb}
\begin{split}
e^0_{\mu}=(N,{\bf 0})\;\;\;,\;\;\;e^a_{\mu}=(N^a,h^a_{\,\,i})\\
e^{\,\,\mu}_{0}=(1/N,-N^i /N)\;\;,\;\;e^{\,\,\mu}_{a}=(0,h_a^{\,\,i}),
\end{split}
\eeqa
where $N^i\equiv h_a^{\,\,i}N^a$ with $h^a_{\,\,i}h_b^{\,\,i}=\delta^a_{\,\,b}$ 
for the induced 3-vierbein $h^a_{\,\,i}$. This choice 
is considered in \cite{gravity and gauge symm} as a time gauge, which shows
a 3+1 decomposition of the spacetime with the time direction
$e^0_{\mu}=(N,{\bf 0})$, coinciding with the unit normal vector $u_\mu=(N,\bf 0)$
of the 3-surface.
For a formal approach in teleparallel gravity, the time gauge is
imposed after the construction of the Hamiltonian formulation which yields
some second-class constraints on the theory~\cite{gravity and gauge symm,H TEGR}. 
However, we apply directly the specific 
vierbein (\ref{ADMvb}) to the quadratic
computation of TEGR given that physical degrees of freedom in addition to GR
will end up redundant as they appear 
merely in the total derivative term as implied in the relation (\ref{T}).

It is straightforward to define the 3-covariant derivative $D_i$ with respect to 
the 3-Weitzenbock connections
$\mbox{}^{(3)}\Gamma^{i}_{\,\,jk}=h_a^{\,\,i}\partial_k h^a_{\,\,j}$, which can 
satisfy the relations
\beq \label{MC}
D_i h_{jk}=0\;\;;\;\;D_i N^j=h_a^{\,\,j}\partial_i N^a,
\eeq
where $h_{ij}$ is the induced 3-metric corresponding to the metric obtained from (\ref{ADMvb}):
\beq \label{ADM}
ds^2=N^2 dt^2-h_{ij}(dx^i+N^i dt)(dx^j+N^j dt).
\eeq
The compatible feature of the  metric is still preserved in the 3-teleparallel geometry due to the first relation in (\ref{MC}).

The non-vanished torsions in this representation are
\begin{eqnarray}
T^{0}_{\,\,j0}&=&\partial_j N/N\,,
\nonumber \\
T^{i}_{\,\,j0}&=&D_j N^i -\frac{N^i}{N}\partial_j N -h_a^{\,\,i} \partial_0 h^a_{\,\,j}\,,
\nonumber \\
T^{i}_{\,\,jk}&=&h_a^{\,\,i}(\partial_j h^a_{\,\,k}-\partial_k h^a_{\,\,j})\equiv \mbox{}^{(3)} T^{i}_{\,\,jk},
\end{eqnarray}
where the last equation is referred to as the definition of the induced 3-torsion.

It is also found to be convenient to define the ``extrinsic torsion" as
\beq
\Sigma_{ij}=\frac{1}{2N}(\dot{h}_{ij}-D_i N_j -D_j N_i)
\eeq
in the following calculations. Note that the extrinsic curvature is given by $\bar{\Sigma}_{ij}=\frac{1}{2N}(\dot{h}_{ij}-\bar{D}_i N_j -\bar{D}_j N_i)$
with $\bar{D}_i$\,the 3-Levi-Civita covariant derivative.
 We can derive the relation between the extrinsic torsion and curvature if further applying the relation (\ref{w-l}) among the induced 3-connections:
\beq
\mbox{}^{(3)}\Gamma^{i}_{\,\,\,jk}=\mbox{}^{(3)}\bar{\Gamma}^{i}_{\,\,\,jk}
+\mbox{}^{(3)}K^{i}_{\,\,\,jk},
\eeq
so that
\beq \label{t-r}
\Sigma_{ij}=\bar{\Sigma}_{ij}-\frac{N^k}{2N}(T_{ijk}+T_{jik}).
\eeq

Under the decomposition (\ref{ADMvb}), the torsion scalar (\ref{T}) is obtained in terms of the extrinsic torsion and 3-torsion as
\beq \label{ADMT1}
T=\Sigma_{ij}\Sigma^{ij}-(\Sigma+\frac{N^k}{N}T^{i}_{\,\,ik})^2-T^{(3)}-2\frac{\partial_kN}{N}T^{i\,\,k}_{\,\,i} +2\frac{N^k}{N} T_{ijk}\Sigma^{ij} +\frac{N^kN^l}{2N^2}T^{ij}_{\;\;k}(T_{ijl}+T_{jil}),
\eeq
where $T^{(3)}=\frac{1}{4}T_l ^{\,\,mn}T^l_{\,\,mn}-\frac{1}{2}T^{mn}_{\,\,\,\,\,\,\;\;l} T^l_{\,\,mn}-T^j_{\,\,jk}T^{l\,\,k}_{\,\,l}$. In order to 
 compare with the usual ADM formalism in GR, we insert the relation (\ref{t-r}) into (\ref{ADMT1}). 
 Using the differential by part $\bar{D}_k(NT^{i\,\,k}_{\,\,i})=N\bar{D}_k T^{i\,\,k}_{\,\,i}+T^{i\,\,k}_{\,\,i}\partial_kN$ 
 for the substitution, we have the final expression, given by
\beq \label{ADMT}
T=\bar{\Sigma}_{ij}\bar{\Sigma}^{ij}-\bar{\Sigma}^2+R^{(3)}+\mathcal{D}_T,
\eeq
with $\mathcal{D}_T=-2\bar{D}_k(NT^{i\,\,k}_{\,\,i})/N$ becoming a total divergence term in the action. 
Here,  $R^{(3)}$ is from the same definition as (\ref{T}) with  $T^{(3)}=R^{(3)}+2\bar{D}_j T^{l\,\,j}_{\,\,l}$.

Replacing $\bar{\Sigma}^{ij}$ by 
$\Pi^{ij}=\sqrt{h}(\bar{\Sigma}^{ij}-\bar{\Sigma}h^{ij})$ 
as the conjugate momentum of $h_{ij}$ \cite{adm}, 
we find that (\ref{ADMT}) is identical to Eq. (10) in \cite{LE GR}, 
which is the 3+1 formulation of TEGR under the Schwinger's 
time gauge consideration \cite{Q GF}. 
As a result, the relation $e^0_{\mu}=u_\mu$ in (\ref{ADMvb}) can be realized 
as locking the time axes of tangent frames to the general coordinate time axis,
while in the teleparallel geometry the time coordinate can be identified definitely 
with $e_A^0$  to be a time-like vector.
Such time gauge, in general, is not necessarily required for 
the 3+1 formulation of TEGR, where the resulting structure can be different from 
the standard ADM formalism~\cite{H TEGR}.

It is easy now to make a comparison with the ADM decomposition of the Ricci scalar $R$~\cite{TMuk}:
\beq
R=\bar{\Sigma}_{ij}\bar{\Sigma}^{ij}-\bar{\Sigma}^2+R^{(3)} +\mathcal{D}_R,
\eeq
where $\mathcal{D}_R=2 \partial_t(\sqrt{h}\bar{\Sigma})/(N\sqrt{h}) -2\bar{D}_i(\bar{\Sigma}N^i + h^{ij}\partial_j N)/N$. We find that the difference between $T$ and $R$ is given by the total divergence $\mathcal{D}_T-\mathcal{D}_R=2\bar{\nabla}^{\nu}T^{\mu}_{\,\,\mu\nu}$.

\subsection{Scalar Perturbations}
We can now rewrite the action (\ref{LT}) via (\ref{ADMT}) as
\beq \label{ADMLT}
S=\frac{1}{2}\int\sqrt{h}\left[ N(\bar{\Sigma}_{ij}\bar{\Sigma}^{ij}-\bar{\Sigma}^2)
+NR^{(3)}+ N^{-1} (\dot{\phi}-N^i\partial_i\phi)^2
-Nh^{ij}\partial_i\phi\partial_j\phi-2NV \right] +\mathcal{D}_T,
\eeq
where $h^a_{\,\,i}$ and $\phi$ play the role of dynamical variables in which 
one can choose a gauge to fix the time and spatial reparametrizations. 
In the following discussion, we will drop
 $\mathcal{D}_T$ in (\ref{ADMLT}) since it is a total divergent term.
We denote the first order quantities, $\zeta$ and $\gamma$, to parametrize the scalar and tensor perturbations, and use the gauge convenient for studying the quadric action as
\beq \label{para}
\delta\phi=0,\;\;\;h^a_{\,\,i}=ae^\zeta(\delta^a_{\,\,i}+\frac{1}{2}\gamma^a_{\,\,i}),
\eeq
where $\gamma^a_{\,\,i}$ in general contains symmetric and antisymmetric parts, $i.e.$, $\gamma^a_{\,\,i}=\mathbf{s}^a_{\,\,i}+\mathbf{a}^a_{\,\,i}$. The antisymmetric part $\mathbf{a}^a_{\,\,i}$ is a distinct feature for tensor perturbations in teleparalelism.
However, it dose not contribute to the quadric calculation in the discussion given later. 
We define that 
$\gamma_{ij}=\eta_{ab}(\delta^a_{\,\,i}\gamma^b_{\,\,j}+\delta^b_{\,\,j}\gamma^a_{\,\,i})/2$ 
with $\eta_{ab}=\mbox{diag}(-1,-1,-1)$ and
$\gamma_{ii}=\partial_i \gamma_{ij}=0$, leading to the induced metric to be 
\beq \label{full3m}
h_{ij}=a^2e^{2\zeta}(\delta_{ij}+\gamma_{ij}+\frac{1}{4}\gamma_{ai}\gamma^a_{\;\,j}).
\eeq

Temporarily, we concentrate on the scalar quantity $\zeta$ and take $\gamma=0$ in 
(\ref{full3m}) to simplify the calculations. With in mind that $N$ and 
$N^i$ (or $N^a$) are as Lagrange multipliers, we perform the variations
$\delta N$ and $\delta N^i=h^i_a \delta N^a$ to (\ref{ADMLT}), and obtain the constraints
\beqa \label{hmc}
&&h^i_a\bar{\nabla}^j[\bar{\Sigma}_{ij}-h_{ij}\bar{\Sigma}]=0\,,
\nonumber \\
&&R^{(3)}-(\bar{\Sigma}_{ij}\bar{\Sigma}^{ij}-\bar{\Sigma}^2)-N^{-2}\dot{\phi}^2-2V=0\,.
\eeqa
By setting that $N^i=\partial_i\psi+N^i_T$ with $\partial_i N^i_T=0$ and $N=1+N_1$
as suggested  in \cite{nonMal}, we derive  the solutions
\beq \label{scalar solutions}
N_1=\frac{\dot{\zeta}}{H},\;\;N^i_T=0,\;\;\psi=-\frac{\zeta}{a^2 H}+\chi,\;\; \partial^2\chi=\frac{\dot{\phi}^2}{2H^2}\dot{\zeta}.
\eeq
After substituting (\ref{scalar solutions}) back to (\ref{ADMLT}), the Lagrangian 
is expressed in terms of $\zeta$ to the second order  
\beqa
S_\zeta &=&\frac{1}{2}\int dtdx^3 \left\{ ae^{\zeta}\left(1+\frac{\dot \zeta}{H}\right)\left[4\partial^2 \zeta -2(\partial\zeta)^2-2a^2 Ve^{2\zeta} \right] 
\right.
\nonumber \\
        &&\left.~~~~~~~~+a^3 e^{3\zeta}\frac{1}{1+\frac{\dot \zeta}{H}}\left[-6(H+\dot \zeta)^2+\dot \phi^2\right]\right\},
\eeqa
which is identical to the expression by using the background equations (\ref{bgeq}):
\beq \label{qs}
S_\zeta =\frac{1}{2}\int dtd^3x a^3 \frac{\dot{\phi}^2}{H^2}\left( \dot{\zeta}^2-a^{-2} (\partial\zeta)^2\right).
\eeq
This final expression for $S_\zeta$ is identical to the result of the theory (\ref{LR}). 

\subsection{Tensor Perturbations}
As demonstrated in \cite{nonMal}, $\gamma^2$  and  scalar-tensor crossing terms appear
only in the cubic calculations of the Lagrangian, which are beyond the main concern of this paper. Hence, we simply set
$\zeta=0$ in (\ref{full3m}) and consider the contribution only from $\gamma^a_{\,\,i}$. 
After inserting it  to (\ref{ADMLT}), the quadratic action becomes
\beq \label{qt}
S_{\gamma}=\frac{1}{8}\int dtd^3x\;a^3\left[ (\dot \gamma_{ij})^2-a^{-2}(\partial_i \gamma_{jk})^2 \right] .
\eeq
Note that, despite the tensor fluctuation $\gamma^a_{\,\,i}$ contains some extra components, 
i.e., the antisymmetric part $\mathbf{a}^a_{\,\,i}$
comparing  to GR. The contribution of $\mathbf{a}^a_{\,\,i}$ does not show up at this level. 
We can define
$\delta_{ia}\mathbf{s}^a_{\,\,j}=\mathbf{s}_{ij}$\,and\;$\delta_{ia}\mathbf{a}^a_{\,\,j}=\mathbf{a}_{ij}$, so that actually
\beq
\gamma_{ij}=\frac{1}{2}\eta_{ab}(\delta^a_{\,\,i}\gamma^b_{\,\,j}+
\delta^b_{\,\,j}\gamma^a_{\,\,i})= \mathbf{s}_{ij}.
\eeq
From the fact that (\ref{qs}) and (\ref{qt}) are the same results as the theory (\ref{LR}), 
we simply conclude that the changing of the dynamical variable from 
$h_{ij}$\,to\;$h^a_{\,\,i}$ does not alter 
the  propagating degrees of freedom in linear perturbations as the extra part $\mathbf{a}^a_{\,\,i}$ drops out in quadratic computations.
This satisfies the consideration of the choice (\ref{ADMvb}) 
and can be seen from the relation (\ref{T}), where the quantities besides the ADM ansatz
are ignored in advance.
 Therefore, we have two tensor modes plus one scalar mode in teleparallel gravity, and the effects from $\mathbf{a}^a_{\,\,i}$ 
 in higher order perturbations are worth for further investigations.

\section{High Order Generalization}

The high order generalization of TEGR was first proposed in \cite{iwi} as an alternative inflation model. As analogy to $f(R)$ gravity, 
the modification of the torsion scalar $T$ can be generalized to an arbitrary function with the action 
\beq \label{Lft}
S=\frac{1}{2}\int dx^4 e f(T).
\eeq 
It is interesting to note that the formalism of \cite{Hehl}
is still applicable to the theory in~(\ref{Lft}) by introducing
a 
constraint directly to the spin-connection instead of the curvature.

In order to study the quadratic action of the high order Lagrangian, 
it is provided to be convenient to perform the conformal transformation 
\cite{TMuk}. We simply review the transformation of $f(T)$ gravity to 
the Einstein frame first with the detail given in \cite{CT}.
We use the formulation
\beq
f(T)=FT-2V,\;\;F\equiv \frac{df}{dT},\;\;V=\frac{FT-f(T)}{2},
\eeq
and define\,$\hat{e}^A_\mu=\sqrt{F}e^A_\mu\equiv\Omega e^A_\mu$. 
The torsion scalar is transformed as
\beq
T=\Omega^2[\hat{T}-4\hat{\partial}^{\mu}\omega\hat{T}^\rho_{\;\,\rho\mu}+6(\hat{\nabla}\omega)^2],
\eeq
where\;$\hat{\partial}^{\mu}\omega\equiv\hat{\partial}^{\mu}\Omega/\Omega$. Since\,$e=\Omega^{-4}\hat{e}$ and $F=\Omega^2$, 
the action (\ref{Lft}) is rewritten as
\beq \label{Eft}
S=\frac{1}{2}\int dx^4\hat{e}\left[\hat{T}-\frac{4}{\sqrt{6}}\hat{\partial}^{\mu}\varphi \hat{T}^\rho_{\;\,\rho\mu} +(\hat{\nabla}\varphi)^2-2U(\varphi)\right],
\eeq
where we have defined that $d\varphi=\sqrt{6}d\omega=\sqrt{6}dF/2F$ and $U=V/F^2$. 
Comparing the action (\ref{Eft})
with  (\ref{LT}), there is an additional scalar-torsion coupling term,
$\hat{\partial}^{\mu}\varphi \hat{T}^\rho_{\;\,\rho\mu}$, 
which breaks the local Lorentz invariance and gives some extra degrees of freedom. 
The corresponding field equation of (\ref{Eft}) 
is obtained by variation with respect to the vierbein field 
$\hat{e}^A_\mu$, where the covariant representation is
\beq \label{efe}
G_{\mu\nu}        =\Theta^{\varphi}_{\mu\nu}+H_{\mu\nu}
\eeq 
with $\Theta^{\varphi}_{\mu\nu}$ the corresponding energy-momentum tensor of the scalar field\;$\varphi$ and 
$H_{\mu\nu}$  obtained from variation of the coupling 
$\hat{e}\hat{\partial}^{\mu}\varphi \hat{T}^\rho_{\;\,\rho\mu}$, given by:
\beqa 
\Theta^{\varphi}_{\mu\nu}&=&\hat{\partial}_\mu \varphi\hat{\partial}_\nu \varphi-
\frac{1}{2}g_{\mu\nu}  (\hat{\nabla}\varphi)^2+g_{\mu\nu}U
\nonumber \\
H_{\mu\nu}               &=&\frac{2}{\sqrt{6}}[\hat{g}_{\mu\nu}\hat{\partial}^\lambda\varphi \hat{T}^\rho_{\;\,\rho\lambda} -\hat{\partial}^\lambda \varphi \hat{T}_{\nu\mu\lambda}-2\hat{\partial}_\nu \varphi\hat{T}^\rho_{\;\,\rho\mu}]-\frac{2}{\sqrt{6}}\hat{e}^{-1}\hat{g}_{\nu\lambda}\hat{e}^A_{\mu} \hat{\partial}_\alpha [\hat{e}(\hat{\partial}^\lambda \varphi e^\alpha_A-\hat{\partial}^\alpha \varphi e^\lambda_A)],\,\
\eeqa
respectively. The variation of the scalar field $\varphi$ gives the equation of motion
\beq
\square\varphi+U^\prime+\frac{2}{\sqrt{6}}\hat{\nabla}^{\mu}\hat{T}^\rho_{\;\,\rho\mu}=0,
\eeq
where $U^\prime\equiv dU/d\varphi$.

It is easy to observe that $H_{\mu\nu}$ in general is not a symmetric tensor, whereas 
$G_{\mu\nu}$ and $\Theta^{\varphi}_{\mu\nu}$ are. 
Therefore, following the discussion in \cite{LV} but here in the Einstein frame, 
the field equation~(\ref{efe}) can be splitted into:
\beqa
G_{\mu\nu}&=&\Theta^{\varphi}_{\mu\nu}+H_{(\mu\nu)}\\
         0&=&H_{[\mu\nu]} \label{extra}
\eeqa
where $H_{\mu\nu}=H_{(\mu\nu)}+H_{[\mu\nu]}$. The presence of (\ref{extra}) makes (\ref{efe}) an equation for all 16 components instead of 10, indicating that the theory (\ref{Eft})  
in fact precesses more degrees of freedoms than the linear teleparallel theory (\ref{LT}). 

Torsion theories with scalar fields have been discussed in \cite{quintaxion},
while the higher order effects of the boundary terms are  considered
in~\cite{Nieh-Yan functional}. Although the generic analysis of the perturbations of 
$f(T)$ theories is virtually important, in the present work we intend to exam practically 
a simplified calculation by specifying the vierbein field $\hat{e}^A_\mu$ in the same 
manner as (\ref{ADMvb}). This vierbein representation reveals the time gauge condition 
from the gauge field approach of quantum gravity \cite{Q GF}. 
The extra degrees of freedom introduced in the
nonlinear teleparallel theories are fixed so that the back ground choice,
$\hat{e}^A_{\mu}=\mbox{diag}(1,\hat{a},\hat{a},\hat{a})$, can be applied directly.

%
 
The torsion scalar $\hat{T}$ is demonstrated by (\ref{ADMT}) where we denote 
$\hat{\Sigma}_{ij}$ 
as the extrinsic curvature with respect to the induced metric $\hat{h}_{ij}$. 
For simplicity, we represent the scalar-torsion coupling under the integration by part as
\beq \label{div lv}
\hat{e}(\varphi \hat{\nabla}^{\mu}\hat{T}^\rho_{\;\,\rho\mu})= -\varphi\partial_t(\sqrt{\hat{h}}\hat{\Sigma})+ \sqrt{\hat{h}}\varphi\bar{D}_i(N^i\hat{\Sigma}+h^{ij}\partial_jN-N\hat{T}^{j\,\,i}_{\,\,j}),
\eeq
where the decomposed $\hat{\nabla}^{\mu}\hat{T}^\rho_{\;\,\rho\mu}$ 
is found in the previous discussion on the total divergence 
$\mathcal{D}_R$ and $\mathcal{D}_T$. Without knowing the background field, 
it is explicit that the term $\sqrt{\hat{h}}\varphi\bar{D}_i(...)$ 
will become a total divergence if we impose the condition $\delta\varphi=0$ 
where $\varphi$ preserves only its background value $\varphi=\varphi(t)$.
This condition is nothing but the unitary field gauge commonly applied for 
computing the primordial curvature perturbations.
The action (\ref{Eft}) under this condition is written as 
\beq \label{ADMft}
S=\frac{1}{2}\int\sqrt{\hat{h}}\left[ N(\hat{\Sigma}_{ij}\hat{\Sigma}^{ij}-\hat{\Sigma}^2)+N\hat{R}^{(3)}+\frac{4}{\sqrt{6}} \dot{\varphi}\hat{\Sigma}+ N^{-1}\dot{\varphi}^2-2NU \right], 
\eeq
where we have neglected two total divergence terms, $\mathcal{D}_{\hat{T}}$ 
from the decomposed $\hat{T}$ and $\sqrt{\hat{h}}\varphi\bar{D}_i(...)$
 as mentioned above.

One may observe that all terms inside (\ref{ADMft}) are no more than variables of the metric $\hat{h}_{ij}$ with its corresponding Levi-Civita derivative $\hat{D}_i$, 
and without higher than second-order time derivative. 
The change of variables from $\hat{h}_{ij}$ to $\hat{h}^a_i$ shows no difference in quadratic computations for $\hat{\Sigma}_{ij}$ and $\hat{R}^{(3)}$ according to the discussion 
in Sec. III, while the only non-metric quantity $\hat{T}^{j\,\,i}_{\,\,j}$ 
drops as the total divergence. Hence, the degrees of freedom of (\ref{ADMft}) 
is the same as the action (\ref{ADMLT}), which is found to have one scalar mode 
and two tenser modes in linear perturbations. 
The theory with the unusual non-minimal derivative coupling between field and gravity 
introduces no new degree of freedom, which is also found in \cite{Germani}.

We can obtain the Hamiltonian and momentum constraints of (\ref{ADMft}) 
from the variation of $N$ and $N^a$, written as
\beqa \label{m and h}
h^j_a\hat{D}^i\left[ \hat{\Sigma}_{ij}-h_{ij}\hat{\Sigma}+h_{ij}\frac{4\dot{\varphi}}{\sqrt{6}N}\right]                                     &=&0, \\
\hat{R}^{(3)}-(\hat{\Sigma}_{ij}\hat{\Sigma}^{ij}-\hat{\Sigma}^2) -N^{-1}\frac{4}{\sqrt{6}}\dot{\varphi}\hat{\Sigma}-N^{-2}\dot{\varphi}^2-2U&=&0,
\eeqa
respectively.
Following the background of the vierbein $\hat{e}^A_{\mu}=\mbox{diag}(1,\hat{a},\hat{a},\hat{a})$,
we are able to use the same parametrization as given in (\ref{para}). 
The background field equations and equation motion of $\varphi$ are  
\beqa \label{bg ft}
3H^2&=&\frac{1}{2}\dot{\varphi}^2+U+\sqrt{6}\dot{\varphi}H 
\nonumber \\
\dot{H}&=&-\frac{1}{2}\dot{\varphi}^2+\frac{1}{\sqrt{6}}\ddot{\varphi}-\frac{\sqrt{6}}{2}H\dot{\varphi}
\nonumber\\
0&=&\ddot{\varphi}+3H\dot{\varphi}+U^\prime+\sqrt{6}(\dot{H}+3H^2), 
\eeqa
where the equation of continuity is satisfied, while the field equations are non-trivially coupled.
We denote the perturbations as $N=1+N_1$ and $N^i=\partial_i\psi+N^i_T$, and
hide the notation of the conformal frame in convenience for the following discussion. 
The quadratic action of (\ref{ADMft}) is given by
\beqa \label{S2 1}
S_\zeta &=&\frac{1}{2}\int dtdx^3 \left\{ ae^{\zeta}\left(1+N_1\right)\left[-4\partial^2 \zeta -2(\partial\zeta)^2-2a^2 Ue^{2\zeta} \right] 
\right.
\nonumber \\
        &&\left.+a^3 e^{3\zeta}\frac{1}{1+N_1}\left[-6(H+\dot \zeta)^2+\dot \phi^2+4(H+\dot \zeta)\partial^2\psi\right]
\right.\nonumber \\
        &&\left.+a^3 e^{3\zeta}\frac{\dot{\varphi}}{1+N_1}\frac{4}{\sqrt{6}}\left[ 3(H+\dot \zeta)-\partial^2\psi\right] 
        \right\}.
\eeqa
After some simple manipulation together with the constraint (\ref{m and h}) and 
 background equations (\ref{bg ft}), we rewrite  (\ref{S2 1})  as
\beqa \label{S2 2}
S_\zeta &=&\frac{1}{2} \int  dtdx^3a^3 \left[ -6\mathcal{G}_T\dot{\zeta}^2-2\mathcal{F}_T (\partial\zeta)^2+2\Xi N_1^2
\right. \nonumber \\        
         &&-\left. 4\Theta N_1 \partial^2\psi+4\mathcal{G}_T\dot{\zeta}\partial^2\psi+12\Theta N_1\dot{\zeta}-4\mathcal{G}_T N_1\partial^2\psi  \right], 
\eeqa
where $\mathcal{G}_T=\mathcal{F}_T=1$ , $\Xi=-U$ and $\Theta=H(1-\dot{\varphi}/\sqrt{6}H)$. 
The action (\ref{S2 2}) is in fact the representative of 
the generic quadratic action with the second-order field equations given in \cite{GG}. 
Following the discussion therein, we can arrive directly at 
the final expressions for tensor and scalar perturbations as
\beqa \label{qft}
S_{\gamma}&=&\frac{1}{8}\int dtd^3x\;a^3\left[ \mathcal{G}_T(\dot \gamma_{ij})^2-a^{-2}\mathcal{F}_T(\partial_i \gamma_{jk})^2 \right] ,\\
S_\zeta   &=&\frac{1}{2}\int dtd^3x a^3 \left[ \mathcal{G}_S\dot{\zeta}^2-a^{-2}\mathcal{F}_S (\partial\zeta)^2\right],
\eeqa
respectively, where
\beqa
\mathcal{F}_S &=& \frac{1}{a}\frac{d}{dt}\left( \frac{a}{\Theta}\mathcal{G}_T^2\right)-\mathcal{F}_T, \\
\mathcal{G}_S &=& \frac{\Xi}{\Theta^2}\mathcal{G}_T^2+3\mathcal{G}_T. 
\eeqa
We find that tensor perturbations are unchanged from the familiar result (\ref{qt}) 
despite the presence of the scalar-torsion coupling in (\ref{ADMft}), while 
the scalar perturbations shall satisfy the conditions of  $\mathcal{F}_S>0$ 
and $\mathcal{G}_S>0$ to avoid ghost and instabilities \cite{GG}
if a certain $f(T)$ inflation model is considered under the time 
gauge condition of the vierbein field.

\section{Conclusions}

In order to study the primordial behavior for gravity constructed under teleparallelism, 
especially for TEGR, we have performed a specific 3+1 decomposition of the vierbein field  
in light of the ADM consideration on perturbations.
The corresponding metric obtained from the concerning vierbien is 
the ADM formulation commonly used in models of inflation. 
We have found that the torsion scalar $T$ under this representation 
differs from the decomposed Ricci scalar $R$ by a total divergence, 
which satisfies the general relation between $T$ and $R$. 
Despite that the local Lorentz invariance is broken by the parallelizable
condition of the theory,
the uncovered (new) physical quantities beyond GR never appear with
dynamical importance in our result even in the extension to $f(T)$ gravity.
This special decomposition can also be interpreted as 
a specific Hamiltonian formulation under the Schwinger's time gauge condition
which fixes the Lorentz violation issue simultaneously.
 
The quadratic computation of $T$ can be taken as a rephrase of the standard result 
by changing the variable from $h_{ij}$ to $h^a_{\,\,i}$. 
The reason is that linear perturbations give identical formulations to metric variables 
while all non-metric terms are removed as total divergences.
Although in general the tensor part of the vierbein contains extra components, 
its effects do not show up at the quadratic level, and the quantization 
process afterword will be no difference from the content of curvature perturbations.

The high order generalization of TEGR has also been considered as an alternative scenario 
to driven inflation under a purely gravitational effect. 
However, to discuss the cosmological perturbations, 
one has to address first the extra degrees of freedom introduced by the parallelizable
topological condition in teleparallel theories. 
We have demonstrated in this work a primary approach on $f(T)$ theories by fixing 
the vierbein field under a time gauge condition, as already applied in TEGR.   
The resulting theory reveals a special kind of curvature perturbation actions with
second order field equations.
Nevertheless, despite the local Lorentz invariance can be brought back to teleparallel
gravity through the time gauge fixing,
those extra degrees of freedom have been found with interesting phenomenological effects during the
evolution of the universe. It is worthy of a future investigation for the general
studies on inflation theories constructed within teleparallelism.

\begin{acknowledgments}
 We are grateful to E. N. Saridakis and T. Kobayashi for the inspiring and helpful discussions. We thank KITPC for the wonderful programs
of ``Dark Matter and New Physics'' and ``String Phenomenology and Cosmology''.
This work was partially supported by National Center of Theoretical
Science and  National Science Council (NSC-98-2112-M-007-008-MY3) of R.O.C.
\end{acknowledgments}



\begin{thebibliography}{99}

\bib{TMuk}
V.~F.~Mukhanov, H.~A.~Feldman and R.~H.~Brandenberger,
Phys. Rep. 215, 203 (1992).

\bib{nonMal}
J.~M.~Maldacena, JHEP 0305, 013 (2003).

\bib{GG}
T.~Kobayashi, M.~Yamaguchi and J.~Yokoyama,
  Prog.\ Theor.\ Phys.\  {\bf 126}, 511 (2011).


\bib{GI}
T. Kobayashi, M. Yamaguchi and J. Yokoyama, Phys. Rev. Lett. 105, 231302 (2010);
X.~Gao, T.~Kobayashi, M.~Yamaguchi and J.~Yokoyama,
  Phys.\ Rev.\ Lett.\  {\bf 107}, 211301 (2011);
  A. D. Felice and S, Tsujikawa, Phys. Rev. D84, 083504 (2011).

\bib{GMG}
A. D. Felice and S, Tsujikawa, JACP 04, 029 (2011).

\bib{Gao}
 X.~Gao,
 JCAP {\bf 1110}, 021 (2011);
 X.~Gao and D.~A.~Steer,
  JCAP {\bf 1112}, 019 (2011).

\bib{Inf_bi}
F.~E.~Schunck, F.~V.~Kusmartsev and E.~W.~Mielke,
  Gen.\ Rel.\ Grav.\  {\bf 37}, 1427 (2005);
 E.~W.~Mielke, F.~V.~Kusmartsev and F.~E.~Schunck, 
 {\it Proceedings of the Eleventh Marcel
Grossmann Meeting on General Relativity},
edited by H. Kleinert, R.T. Jantzen and R. Ruffini,
 (World Scientific, Singapore 2008), p. 824-843.

\bib{iwi}
R. Ferraro and F. Fiorini, Phys. Rev. D75, 084031 (2007);
 Phys. Rev. D78, 124019 (2008).

\bib{ft1}
G.~R.~Bengochea and R.~Ferraro,
  Phys.\ Rev.\ D \textbf{79}, 124019 (2009).

  \bib{ft2}
  E.~V.~Linder,
  Phys.\ Rev.\ D \textbf{81}, 127301 (2010);  K.~Bamba, C.~Q.~Geng and C.~C.~Lee,
  arXiv:1008.4036 [astro-ph.CO].

 
   \bib{ft3}
       P.~Wu and H.~W.~Yu,
  Phys.\ Lett.\ \textbf{B692}, 176 (2010);
  Phys.\ Lett.\ \textbf{B693}, 415 (2010);
  Eur.\ Phys.\ J.\ \textbf{C71}, 1552 (2011);
  Phys.\ Lett.\  {\bf B703}, 223 (2011).

   \bib{ft4}
 R.~Myrzakulov,
  Eur.\ Phys.\ J.\ C {\bf 71}, 1752 (2011);
  K.~K.~Yerzhanov, S.~R.~Myrzakul, I.~I.~Kulnazarov and
R.~Myrzakulov,
  arXiv:1006.3879 [gr-qc].

 \bib{ft5}
  R.~Zheng and Q.~G.~Huang,
  JCAP \textbf{1103}, 002 (2011).

 \bib{ft6}
  K.~Bamba, C.~Q.~Geng, C.~C.~Lee and  L.~W.~Luo,
  JCAP \textbf{1101}, 021 (2011).

 \bib{ft7}
    T.~Wang,
  Phys.\ Rev.\  {\bf D84}, 024042 (2011).

 
 \bib{ft9}
    G.~R.~Bengochea,
  Phys.\ Lett.\  {\bf B695}, 405 (2011);

\bib{ft10}
S.~H.~Chen, J.~B.~Dent, S.~Dutta and E.~N.~Saridakis,
  Phys.\ Rev.\ D \textbf{83}, 023508 (2011);
J.~B.~Dent, S.~Dutta, E.~N.~Saridakis,
  JCAP {\bf 1101}, 009 (2011).


\bib{ft11}
 X.~c.~Ao, X.~z.~Li and P.~Xi,
  Phys.\ Lett.\  {\bf B694}, 186 (2010).

\bib{ft12}
   Y.~Zhang, H.~Li, Y.~Gong and Z.~H.~Zhu,
  JCAP {\bf 1107}, 015 (2011).


  \bib{ft13}
  Y.~-F.~Cai, S.~-H.~Chen, J.~B.~Dent, S.~Dutta and E.~N.~Saridakis,
  Class.\ Quant.\ Grav.\  {\bf 28}, 215011 (2011).
    
  \bib{ft14}
   R.~Ferraro and  F.~Fiorini,
  Phys.\ Lett.\  {\bf B702}, 75 (2011);
  Int.\ J.\ Mod.\ Phys.\ Conf.\ Ser.\  {\bf 3}, 227 (2011).


  \bib{ft15}
   S.~Chattopadhyay and U.~Debnath,
  Int.\ J.\ Mod.\ Phys.\  {\bf D20}, 1135 (2011).

 \bib{ft16}
  M.~Sharif and S.~Rani,
  Mod.\ Phys.\ Lett.\  {\bf A26}, 1657 (2011).

 \bib{ft17}
   H.~Wei, X.~P.~Ma and H.~Y.~Qi,
  Phys.\ Lett.\  {\bf B703}, 74 (2011).
  
 \bib{ft18}
   R.~X.~Miao, M.~Li and Y.~G.~Miao,
  JCAP {\bf 1111}, 033 (2011).

 \bib{ft19}
  C.~G.~Boehmer, A.~Mussa and N.~Tamanini,
  Class.\ Quant.\ Grav.\  {\bf 28}, 245020 (2011).
 
  \bib{ft20}
    H.~Wei, H.~Y.~Qi, X.~P.~Ma,
  arXiv:1108.0859 [gr-qc];
   H.~Wei,
  Phys.\ Lett.\ B {\bf 712}, 430 (2012).
  
  \bib{ft21}
  S.~Capozziello, V.~F.~Cardone, H.~Farajollahi and A.~Ravanpak,
  Phys.\ Rev.\ D {\bf 84}, 043527 (2011).
  
  
  \bib{ft22}
  M.~H.~Daouda, M.~E.~Rodrigues and M.~J.~S.~Houndjo,
  Eur.\ Phys.\ J.\ C {\bf 72}, 1890 (2012).
  
     
  \bib{ft23}
  C.~Q.~Geng, C.~C.~Lee, E.~N.~Saridakis and Y.~P.~Wu,
 Phys. \ Lett.\  {\bf B704}, 384 (2011);
C.~Q.~Geng, C.~C.~Lee and E.~N.~Saridakis,
 JCAP {\bf 1201}, 002 (2012).


 \bib{ft24}
  K.~Bamba and C.~Q.~Geng,
   JCAP {\bf 1111}, 008 (2011).

\bib{LV}
B.~Li, T.~P.~Sotiriou and J.~D.~Barrow, Phys. Rev. D83, 064035 (2011).

\bib{dof}
M.~Li, R.~X.~Miao and Y.~G.~Miao, JHEP. 1107, 108 (2011).

\bib{CT}
R.~J.~Yang, Europhys. Lett. 93, 60001 (2011).

\bib{LSS}
B.~Li, T.~P.~Sotiriou and J.~D.~Barrow, Phys. Rev. D83, 104017 (2011).

\bib{E}
A. Einstein, Sitz. Preuss. Akad. Wiss., 1928, p. 217;
A. Einstein, Sitz. Preuss. Akad. Wiss., 1928, p. 224;
A. Unzicker, T. Case, [arXiv:physics/0503046].

\bib{NGR}
K. Hayashi and T. Shirafuji, Phys. Rev. D19, 3524 (1979); 
 Phys. Rev. D24, 3312 (1982), Addendum.


\bib{T intro}
R.~Aldrovandi and J.~G.~Pereira, An Introduction to
Teleparallel Gravity, Instituto de Fisica Teorica, UNSEP,
Sao Paulo (http://www.ift.unesp.br/gcg/tele.pdf).

\bib{gravity and gauge symm}
M. Blagojevic, Gravitation and Gauge Symmetries, Institute of
Physics Publishing, Bristol, 2001. 

\bib{Hehl}
F. W. Hehl {\it et al.},  Phys. Rep. {\bf 258}, 1 (1995).


\bib{HT}
J. W. Maluf, J. Math. Phys. 35, 335 (1994).

\bib{A.A}
A. A. Starobinsky, Phys. Lett. B91, 99 (1980)

\bib{Ortin}
T. Ortin, \textit{Gravity and Strings}, Cambridge University Press, 2004.

\bib{Weinberg}
S. Weinberg, \textit{Gravitation and Cosmology: Principles and 
Applications of the General Theory of Relativity}, John Wiley
\& Sons, New York, 1972.




\bib{LE GR}
J. W. Maluf, J. Math. Phys. 36, 4242~(1995) .

\bib{Ann phys}
E.W. Mielke, Ann. Phys. (N.Y.) {\bf 219}, 78 (1992).



\bib{H TEGR}
J. W. Maluf and J. F. da Rocha-Neto, Phys. Rev. D64, 084014 (2001).

\bib{adm}
R. Arnowitt, S. Deser, and C. W. Misner, in Gravitation: An
Introduction to Current Research, edited by L. Witten (Wiley,
New York, 1962).


\bib{Q GF}
J. Schwinger, Phys. Rev. 130, 1253~(1963).






\bib{quintaxion}
E.~W. Mielke, E.~S.~Romero, Phys. Rev. D {\bf 73}, 043521 (2006).

\bib{Nieh-Yan functional}
E.~W. Mielke, Phys. Rev. D{\bf 80} 067502 (2009); see also  D.~Kreimer and E.~W.~Mielke,
  Phys.\ Rev.\ D{\bf 63}, 048501 (2001);
  E.~W. Mielke,  Nucl.\ Phys.\ B{\bf 622}, 457 (2002).
 

\bib{Germani}
C. Germani and A. Kehagias, Phys. Rev. Lett. 105, 011302 (2010);
JCAP 05, 019 (2010).

\end{thebibliography}
\end{document}